# Symmetry-Breaking Polymorphous Descriptions for Correlated Materials without Interelectronic $U$

Yubo Zhang,[1,3] James Furness,[1] Ruiqi Zhang,[1] Zhi Wang,[2] Alex Zunger,[2,*] Jianwei Sun[1,*]

[1]Department of Physics and Engineering Physics, Tulane University, New Orleans, LA 70118, USA
[2]Renewable and Sustainable Energy Institute, University of Colorado, Boulder, Colorado 80309, USA
[3]Department of Physics, Southern University of Science and Technology of China, Shenzhen 518055, China

Corresponding authors: alex.zunger@colorado.edu and jsun@tulane.edu

Correlated materials with open-shell $d$- and $f$-ions having degenerate band edge states show a rich variety of interesting properties ranging from metal-insulator transition to unconventional superconductivity. The textbook view for the electronic structure of these materials is that mean-field approaches are inappropriate, as the interelectronic interaction $U$ is required to open a band gap between the occupied and unoccupied degenerate states while retaining symmetry. We show that the latter scenario often defining what Mott insulators are, is in fact not needed for the 3d binary oxides MnO, FeO, CoO, and NiO. The mean-field band theory can indeed lift such degeneracies in the binaries when nontrivial unit cell representations (polymorphous networks) are allowed to break symmetries, in conjunction with a recently developed non-empirical exchange and correlation density-functional without an on-site interelectronic interaction $U$. We explain how density-functional theory (DFT) in the polymorphous representation achieves band gap opening in correlated materials through a separate mechanism to the Mott-Hubbard approach. We show the method predicts magnetic moments and gaps for the four binary monoxides in both the antiferromagnetic and paramagnetic phases, offering an effective alternative to symmetry-conserving approaches for studying a range of functionalities in open $d$- and $f$-shell complex materials.

## I. Introduction

Studies of late 3$d$ transition-metal monoxides (MnO, FeO, CoO, and NiO) and their transport properties led to the seminal concepts of Mott insulators and strong correlation.[1] Historically, these discussions were centered around experiments on NiO, a transparent and magnetic insulator. Assuming a non-magnetic (NM) configuration and that the macroscopically observed global cubic symmetry with a single formula unit can be interpreted also locally on an atom by atom basis, early naïve band theory incorrectly predicted NiO to be metallic with partially filled $d$-bands.[2] Later, when low-tempserature antiferromagnetic (AFM) order was considered by doubling the unit cell and allowing magnetic moment formation, the band gap opened up, even in simple band theory.[3] But this triumph of mean field theory to explain band gap opening in the spin ordered phase was considered to be insufficient to explain the observed band gap in the high temperature spin disordered paramagnetic (PM) phase. This is because when a single formula unit per cell was used to describe the PM phase, the local magnetic moment at the transition-metal site must coincide with the global magnetic moment, which is zero in a cell containing a single 3$d$ ion. Thus, the PM phase has to be non-magnetic and hence gapless from band theory, in contrast to experiments on the binary 3$d$ oxides.

This historic failure of such simple band theory ideas set the stage for alternative strongly correlated strategies. In his seminal work,[1] Nevill Mott theorized that the insulating behavior of 3$d$ transition-metal oxides can emerge from the strong correlation, encoded by the on-site inter-electronic repulsion ("$U$") between $d$ electrons. This repulsion keeps the $d$ electrons localized within bands of width $W$ where $U > W$, and was argued as the correct mechanism regardless of the magnetic order.[1] This picture of d-orbital dominated doubly occupied and empty band edges, (valence and conduction bands, respectively), is the textbook model of (Mott) insulation. Many contemporary non-perturbative methods[4-6] applicable to open-shell transition-metal and rare-earth compounds are rooted in the concept of such symmetry conserving approaches, i.e. modeling strong correlation while conserving the unbroken spatial unit cell symmetry.[7,8] This textbook view of strong correlation disqualified both mean-field theories and methods based on perturbation theories for the BO monoxides as well as ternary $ABO_3$ oxides (B=3$d$) band gap problem.

While the Mott-Hubbard Hamiltonian approach seemed *a priori* plausible and potentially exact, it was not obvious if the celebrated Mott Insulator *compounds* –the binary and ternary 3d oxides—follow the edicts of this Hamiltonian, e.g. have 3d-like band edge states with band gaps that essentially equal U. On the other hand, the complexity of the explicitly dynamically correlated methods makes electronic structure studies of real-life structurally complex transition-metal oxides difficult if not intractable for numerous material specific properties in applications such as catalysis, superconductivity, magnetoresistance, and carrier doping.

That the assumption underlying the early naïve band theory[9,7,10,11] that required all transition-metal sites in a PM phase to be symmetry equivalent may be overly restrictive was noted by Trimarchi *et al*[12] and Varignon *et al*.[13] Indeed, much like the AFM configuration that involves a doubling (or quadrupling) of the primitive cell, the PM phase can also be described by an even larger spin supercell. In this larger spin supercell the total magnetic moment is zero as required by paramagnetism, yet local sites can develop nonzero local magnetic moments.[12] Such representations for the PM phase allow the existence of a *distribution of different local environments* (a "polymorphous network") for transition-metal sites.[12] This generalization of mean field for allowing symmetry breaking has produced finite band gaps, local atomic displacements, and moments in both BO binary[12] and $ABO_3$ ternary[13,14] oxides without recourse to explicitly including dynamic correlation.

The success of the polymorphous representations[12-14] relied on several key factors: (i) the use of larger-than-primitive unit cells does not force upon us the symmetry equivalence of all chemically identical sites, thus the polymorphous representation provides an option to break the local symmetry, should this lower the total energy. In other words, although unlike the AFM phase the PM phase has no gap-opening long-range order (LRO), this does not preclude the latter from having gap-opening short-range order (SRO; not disorder). (ii) The use of an exchange-correlation functional in DFT that can distinguish occupied from unoccupied orbitals and allow orbitals to be spatially compact so as to benefit from these symmetry-breaking energy-lowering opportunities, and (iii) "nudging" of the system to allow relaxation in all degrees of freedom, both local displacements and breaking the orbital occupation pattern, thus allowing factors (i) and (ii) to develop. Polymorphous representations[12] allow for a *range* of symmetry breaking mechanisms such as Jahn-Teller displacements, octahedral tilting and rotations, different spin-disordered environments, as well as unequal occupation of previously degenerate states [e.g., for doubly degenerate E level with occupations E($x,y$), we use $E(1,0)$ instead of $E(\frac{1}{2};\frac{1}{2})$]. These broken symmetries can all contribute to band gap opening, magnetic moment formation, and stabilization of the observed crystal structures.[12] Symmetry can be restored as a follow-up step,[15,16] but this formality should have a negligible effect on the total energy when the individual symmetry-broken configurations are spatially localized and have band gaps.

Not all exchange-correlation functionals can take advantage of the energy-lowering symmetry-breaking opportunities afforded by polymorphous representations (obviously, highly-delocalized orbitals produced by some XC functionals may be too "far sighted" to "see" local symmetry breaking[17]). Benefiting from symmetry breaking requires[12-14] density-functionals that have (i) reduced self-interaction error (SIE) leading to realistically compact orbitals, and (ii) the capability to distinguish occupied from unoccupied orbitals[12-14] for example through different effective potentials for different orbitals. Such orbital specific effective potentials can be achieved through explicit orbital dependence, at the 4th and 5th level of the Perdew-Schmidt hierarchy of XC functionals,[18] or implicitly through the kinetic energy density at the 3rd level as indicated in Ref [14]. (Note that Refs [12,13] suggested the stronger restriction of using rung 4 or 5 functionals, but the current work as well as Ref [14] find that for band gap calculations the weaker conditions of Rung 3 can suffice.) When combined with polymorphous representations such density-functionals are far simpler than the explicitly correlated approaches applicable to solids such as Quantum Monte Carlo,[6] dynamical mean field theory (DMFT),[4] and density matrix embedding theory.[5] Moreover, this approach points to a different *mechanism of gap opening relative to the Mott*-Hubbard scenario.

DFT+$U$[19] satisfies the above criteria and thus opens band gaps for open-shell transition-metal compounds.[12,13] It remains non-trivial to determine the $U$ value as an input however, a problem shared with DMFT. The appearance of "+$U$" in DFT+$U$ can create the impression that the success of this method in conjunction with the polymorphous representation in explaining much of the phenomenology of the AFM and PM phases is due to the Mott-Hubbard like correlation physics. However, $U$ appears in DFT+$U$ predominantly as a simplified self-interaction-error (SIE) reduction device that enhances the spatial compactness of 3$d$ orbitals, allowing them to take advantage of symmetry breaking. Furthermore, it has been shown in Ref. [14] that even without $U$, the non-empirical *strongly constrained and appropriately normed* (SCAN)[20] density-functional in conjunction with polymorphous representations opens band gaps of the ternary ABO$_3$ perovskites when symmetry breaking is allowed. The success of DFT-without-$U$ indicates that the Mott mechanism (where the gap equals $U$) does not necessarily apply to the main 3$d$ oxides.

This recent development raises a very important question: *why* can DFT without $U$ open band gaps for correlated materials in both AFM and PM phases of binary transition-metal monoxides and ternary ABO$_3$? The present paper aims to answer this question and apply SCAN for the first time to the prototypical binary transition-metal monoxides.

This paper is organized as following. Section II gives the computational details. In section III, we show computationally that without invoking $U$, the SCAN functional in conjunction with polymorphous representations systematically predicts magnetic moments and qualitatively opens band gaps in four monoxides (MnO, FeO, CoO, and NiO) in both the AFM and PM phases. In the high-temperature PM phase the polymorphous network of spin disorder is modeled by the special quasirandom structure (SQS).[20,21,22] In section IV, we provide an understanding of how the choice of exchange correlation functionals with the right attributes allows band gap opening for correlated materials in conjunction with symmetry breaking representations. The understanding is illustrated in section V by showing results including (i) the success of single determinant SCAN for reproducing the total energies of hydrogen systems from highly correlated methods, (ii) the improved compliance of SCAN with the generalized Koopmans condition, and (iii) the achievement of more spatially compact orbitals with SCAN enabling more effectively symmetry breaking. Finally, Section VI discusses the role of the polymorphous representation and Sec VII provides the summary and conclusions.

## II. Computational Details

The SCAN functional is implemented in the Vienna Ab-initio Simulation Package (VASP)[23] and Turbomole.[24] For the binary systems we used the projector-augmented wave method,[25,26] and a cutoff energy of 500 eV to truncate the plane waves. The $K$-meshes for Brillouin zone integration are 8×8×8 for the 4-atom unit-cell, 3×3×3 for the 64-atom supercell, and 2×2×2 for the 216-atom supercell. We used here projector-augmented waves (PAW) generated from the PBE functional, as SCAN optimized PAWs were not available. It has been shown that using PBE PAWs for other functionals has little influence on calculated energy differences.[27] The hydrogen chain at the thermodynamic limit was modeled in a supercell with the chain along the x direction, while the lattice constants in the y- and z-directions were set to 30 Å, in order to avoid the interaction between two adjacent periodic images. For the total energy calculations of the chain, energies were converged to within $10^{-6}$ eV and a plane-wave energy cut-off of 1200 eV were used.

**Table I.** Stabilities, band gaps, and local magnetic moments of four 3$d$ monoxides in the G-type AFM phase (where every nearest neighbor pair of magnetic moments are antiparallel) and the PM phase calculated by the SCAN density-functional without $U$. The PM phases are modeled by the special quasirandom structure (SQS) with 216-atom supercells to create close approximations to the random spin configuration.[20,12] "Unrelaxed" means that both the lattice type and the cell internal coordinates are kept equal to the experimental NaCl crystal structure. "Relaxed" means that our calculations started from the NaCl experimental crystal structure and used a gradient relaxation algorithm to identify the nearest minima, generally not the deepest minima. The magnetic moments in the SQS-PM phase have small variations in magnitude among the transition-metal sites, and the results given here are the averaged values. See calculation details at the end of the main text.

| Structure & model | Energy (meV/atom) | | | | Band gap (eV) | | | | Magnetic moment (μB) | | | |
|---|---|---|---|---|---|---|---|---|---|---|---|---|
| | MnO | NiO | FeO | CoO | MnO | NiO | FeO | CoO | MnO | NiO | FeO | CoO |
| NM, Unrelaxed | 1487 | 666 | 561 | 542 | 0 | 0 | 0.71[a] | 0 | 0 | 0 | 0 | 0 |
| AFM, Unrelaxed | 0 | 0 | 0 | 0 | 1.63 | 2.48 | 0.22 | 0.98 | 4.44 | 1.58 | 3.55 | 2.58 |
| AFM, Relaxed | −4.7 | −0.2 | −59 | −1.5 | 1.67 | 2.52 | 0.14 | 0.94 | 4.43 | 1.57 | 3.54 | 2.57 |
| AFM, Expt. | -- | -- | -- | -- | 3.5[28] | 3.5[28] | 2.1[28] | 2.8[28] | 4.58[29] | 1.9[29] | 4.0[29] | 3.8~3.98[29] |
| PM, Unrelaxed | 14 | 40 | 13 | 33 | 0.77 | 1.50 | 0.18 | 0.94 | 4.47 | 1.63 | 3.60 | 2.62 |
| PM, Relaxed | 12 | 37 | −32 | 32 | 0.80 | 1.49 | 0.21 | 1.07 | 4.47 | 1.63 | 3.58 | 2.62 |
| PM, Expt. | -- | -- | -- | -- | 3.7[30] | 4.1[30] | 2.5[30] | 2.4[30] | | | | |

[a]FeO can also have a band gap in its non-magnetic phase, but its total energy is high in comparison with the G-type AFM phase.

Finite system (molecular) calculations were carried out in Turbomole. Hydrogen systems were obtained by extrapolating to the basis set limit from the cc-pVXZ (X = 2, 3, 4, 5) basis sets, as described in Ref [31]. The Mn ion with fractional number of electrons was calculated in the def2-TZVP basis set.[32] Expanded quadrature grids were used in all Turbomole DFT calculations using the control option radsize = 100 to augment the standard level seven Turbomole grid with additional radial points. Spin-symmetry was explicitly broken in initial guess orbitals.

Supercells modelling a PM phase with a given lattice symmetry (cubic, orthorhombic, etc.)[12,13] are chosen as follows.
(i) The global shape of a supercell is fixed to the macroscopically observed lattice symmetry.
(ii) Lattice sites of an $N$-atom supercell are occupied by spin-up and spin-down so as to achieve the closest simulation of a perfectly random (i.e. high-temperature limit) distribution, modeled by SQS. The latter selects a supercell of finite number $N$ of atoms so that the spin-spin pair correlation functions best mimic the correlation functions of an infinite sized supercell.[20] Then, $N$ is increased as a convergence parameter until no further change occurs. We use here the 216-atom supercell for the SQS PM. SQS can be generated for random spin-spin correlations (no short-range order at the high-$T$ limit) or for atomic arrangements with some atom-atom correlations (corresponding to the low-$T$ limit of PM).[21] In the current study we consider the fully random PM phase, which contains the essential physics. Agreement with experiment can presumably be improved by including finite-temperature short-range order.

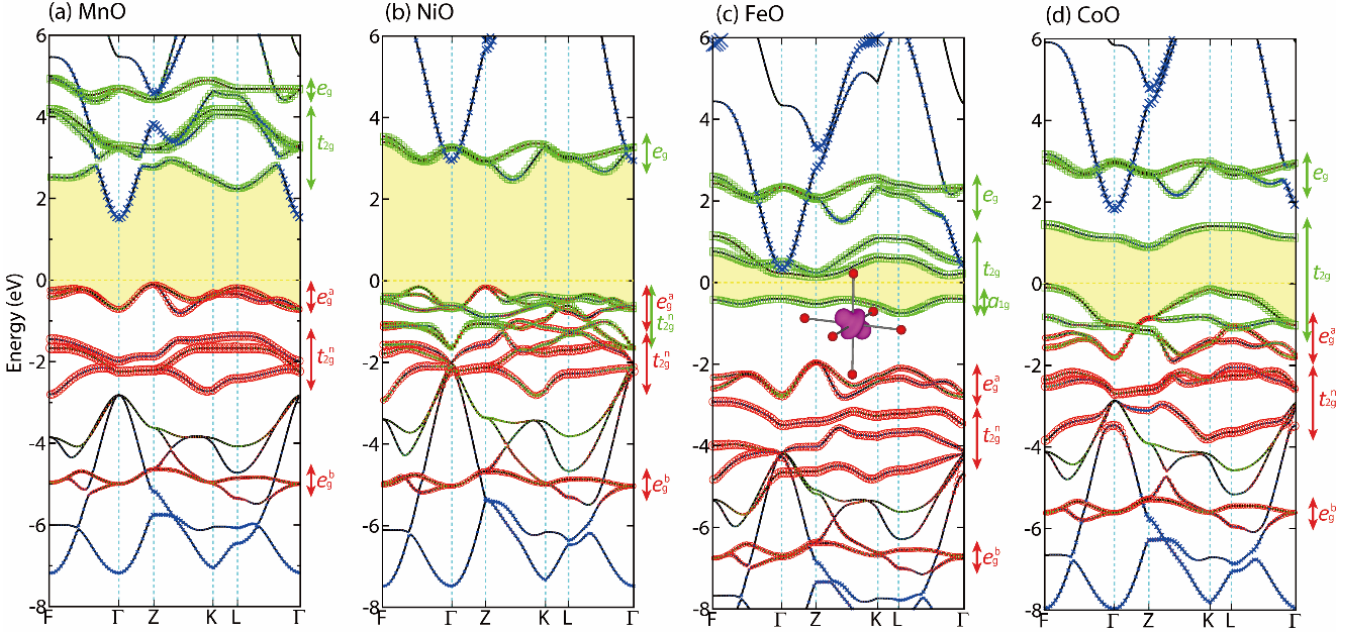

**Figure 1.** SCAN Band structures of the four G-type AFM transition-metal monoxides calculated with the experimental NaCl crystal structures. Orbital characters are indicated where: red circles (○) are the $3d$ states in the majority spin channel, green squares (□) the $3d$ states in the minority spin channel, and blue crosses (×) the transition-metal $4s$ sates. Interaction patterns of the $3d$ states, i.e., bonding ($^b$), anti-bonding ($^a$), and non-bonding ($^n$), are also labeled. For FeO, the inset shows the charge density distribution of the highest occupied band $a_{1g}$ band, which is a linear combination of the $d_{xy}$, $d_{yz}$, and $d_{xz}$ orbitals.[33] The valence band maximum (VBM) is located at the "M" point, which is not included in the standard $K$-path of this figure. The regions between occupied and unoccupied states are shaded with yellow. Calculation details are given at end of the main text.

(iii) Relaxation is performed by retaining the symmetry of the lattice vectors to the originally assumed symmetry (here, cubic) while relaxing cell-internal atomic positions. Atoms can be nudged initially to avoid trapping in local minima.
(iv) Occupation numbers of degenerate partially occupied orbitals are nudged and not forced to be the same.
(v) Wavefunctions are not symmetrized.

We use the conventional two-atom primitive cell for the NM, and the four-atom supercell for the G-type AFM. Note, a 64-atom supercell of FeO is also used for the AFM phase. We used the SQS of Ref 12 for the 64- and 216-atom supercells to model PM phases, which has been obtained via the stochastic generation algorithm implemented in the Alloy Theoretic Automated Toolkit (ATAT) code[34,35].

### III. Results

### A. General trends

Table I shows that SCAN in conjunction with polymorphous representation that permits energy-lowering symmetry-breaking produces the following results:

(i) We predict the naïve NM model to be a high-energy state thus energetically irrelevant to the ground state physics at hand. Thus, the NM calculations previously used extensively in conjunction of DMFT publications (listed in Ref 13) to disqualify DFT are not pertinent.

(ii) Structural relaxation enhances the stability of the monoxides in the AFM and PM phases. This change is small for MnO, NiO, and CoO, while the energy of FeO is significantly lowered by structure relaxation resulting in a structural phase transition to the monoclinic phase. The following discussions are based on calculations from the experimental NaCl crystal structures.

(iii) Our method stabilizes the magnetic moments of the transition-metal sites in both AFM and PM phases with values in good agreement with those measured from their AFM phases. Such findings are consistent with previous work in cuprates[36-38] and the $ABO_3$ perovskites,[14] for which SCAN also predicts reliable magnetic moments.

(iv) Our approach opens the band gaps for both the AFM and PM phases. Furthermore, the orbital-decomposed density of state (DOS) (Figure 2) shows that the VBM of these compounds is far from being a $d$ state as envisioned in the Mott-Hubbard model (that treats only $d$ states). This indicates again that the Mott-Hubbard mechanism involving the interelectronic interaction $U$ is not necessarily the required description for opening band gaps in the classic (so-called) Mott insulators binary (present work) or ternary $3d$ oxide Mott insulators.[13,14] Indeed, unlike the Mott mechanism (gap equals $U$) the mechanism is free from $U$, thereby constituting different physics.

Much like other popular density-functionals however, SCAN produces too small band gaps for the usual (nominally uncorrelated) semiconductors such as Si and GaAs, here we also face the same difficulty. SCAN significantly underestimates the band gap magnitudes, in particular, the gaps for the AFM (PM) phase of FeO is 0.22 (0.18) eV, significantly smaller than the 2.5 (2.5) eV observed experimentally. When compared to the metallic behavior predicted by the popular Perdew-Burke-Ernzerhof (PBE)[39] density functional however, the gap opening is a remarkable step forward. The remaining significant underestimation from SCAN is likely due to the residual SIE. More realistic band gaps were opened by PBE+$U$ in Ref 12, at the price of introducing $U$ as a way to remove SIE within DFT.[40]

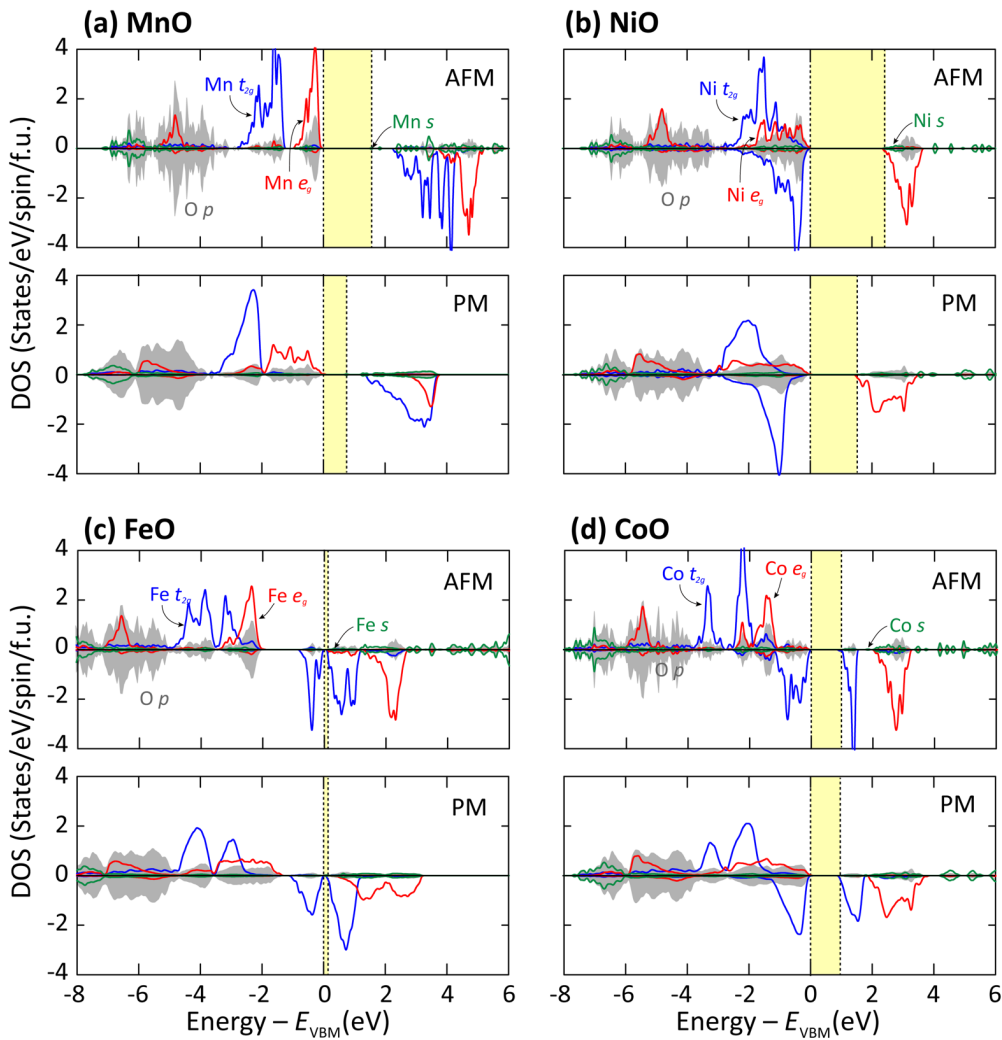

**Figure 2.** Density of states of four transition-metal monoxides with the G-type AFM and SQS-PM spin configurations calculated by SCAN. All crystal structures are fixed to the experimental data ('Unrelaxed' in Table I). The SQS model contains 216 atoms. The band gaps are indicated by the vertical dashed lines. Calculation details are shown at the end of the main text.

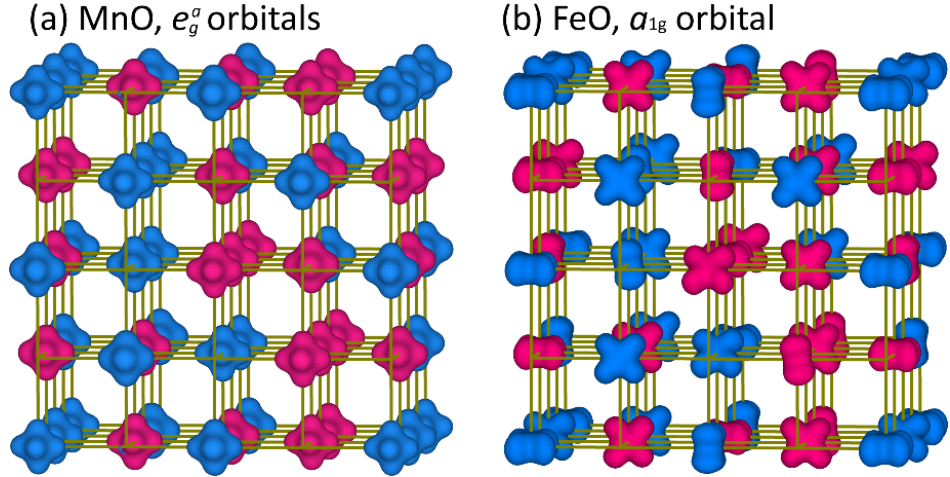

**Figure 3.** Spin and orbital disorder patterns in the paramagnetic phase simulated in the 64-atom-supercell from the SCAN SQS-PM model. (a) MnO, spin disorder visualized from the charge density distributions of the two $e_g^a$ bands. (b) FeO, the $a_{1g}$ band. The two $e_g^a$ bands in MnO and the $a_{1g}$ band in FeO are selected because they are just below the valence band maximum and are well separated from the other transition-metal $3d$ and O-$2p$ states. Similar information for NiO and CoO is difficult to extract because of the significant orbital mixing. The blue and red isosurfaces denote the two spin channels on each transition-metal ion. Oxygen atoms are omitted for simplicity. The 216-atom supercell contains similar information, but the smaller 64-atom supercell is selected here for clarity.

### B. The AFM Phases

The gap opening mechanism of the AFM phase monoxides has been widely studied using different methods including DFT+$U$, see Ref. 12 and references therein for discussions.[33,29,12] Table I shows that the present approach of SCAN, in conjunction with the polymorphous representation without invoking the interelectronic $U$, opens the band gap of the AFM phase for the four transition-metal monoxides, while Figure 1 shows the SCAN band structures. The gap opening mechanism varies among the 4 binary oxides studied here:

(i) In MnO a band gap of 1.63 eV is seen due to exchange splitting between the spin-up and spin-down channels, resulting in the $Mn^{2+}$ $3d^5$ orbitals having one fully occupied [$(t_{2g}^3 e_g^2)^\uparrow$] group and one completely empty [$(t_{2g}^0 e_g^0)^\downarrow$] group ($\uparrow$ and $\downarrow$ denote the spin-majority and spin-minority channels).

(ii) For NiO, the band gap of 2.48 eV is between $(t_{2g}^3)^\downarrow$ and $(e_g^0)^\downarrow$ states, originating from crystal-field splitting.

(iii) For FeO, the band gap is within the three $t_{2g}$ orbitals of $Fe^{2+}$ $3d^6$. If we define the local Cartesian axes along the Fe-O bond directions, the self consistent occupation of $d_{xy}$, $d_{yz}$, and $d_{xz}$ orbitals are 0.332, 0.342, and 0.272 electrons, respectively, demonstrating the occupation symmetry breaking (i.e., orbital anisotropy[29] or polarization[33]). Such occupation symmetry breaking is crucial for the band gap formation in partially occupied $t_{2g}$ systems like FeO ($t_{2g}^1$), in addition to the exchange splitting and crystal-field splitting mechanisms (see further discussions in Appendix A). The occupied minority spin $d$ band is a linear combination of the $d_{xy}$, $d_{yz}$, and $d_{xz}$ orbitals with $a_{1g}$ symmetry.[33]

(iv) For CoO, the band gap opening is due to the occupation polarization of the two occupied $t_{2g}$ sub-bands and one empty $t_{2g}$ sub-band. These findings are consistent with earlier work.[41,40,12]

### C. The PM phases

Table I shows that the present method with the SQS-PM model stabilizes local magnetic moments for all monoxides considered. The values are almost the same as those of the AFM phases, and thus the band gaps are opened. This stabilization of local magnetic moments makes the SQS-PM model substantially more stable than the naïve NM model even though both models have zero total magnetic moment.

In addition to lifting degeneracy, the low-symmetry crystal field due to the short-range spin order in PM broadens DOS in comparison to the AFM DOS for all monoxides considered, as shown in Figure 2. This is consistent with early work for the PM phase.[42,30] As a result, the metal-oxygen bonding is weakened, leading to noticeable destabilizations of the PM monoxides with respect to their AFM phases. The broadening of DOS also causes the reduction of band gaps in MnO (from 1.63 eV in the AFM phase to 0.77 eV in PM) and NiO (from 2.48 eV in the AFM phase to 1.50 eV in PM). Note that we used in the present SQS the high temperature spin arrangement and no attempt was made to introduce short range spatial correlations between the spins which could affect the gap. It is interesting to note that the band gaps of FeO and CoO are almost unaffected in the PM phase. This is likely due to the fact that the gap opening in these two monoxides mainly comes from the occupation symmetry breaking of the $t_{2g}$ states, which is not directly related to the crystal field symmetry breaking.

The DOS broadening highlights the capability of the SQS-PM model to provide different local environments[12] and SCAN's capability to recognize chemical environments.[43] This enabling and appropriate handling of *different local environments* is directly visualized in the density distributions of $3d$ spin orbitals. Figure 3(a) plots the spin orbital density of the two Mn$^{2+}$ $e_g^a$ states below the Fermi level of the SQS-PM MnO [counterparts to the two $e_g^a$ states indicated in Figure 1(a)]. The distribution pattern clearly reflects the random distribution patterns of the magnetic moment directions, while the spatial distributions at Mn$^{2+}$ sites with different magnetic moment directions are equivalent. In stark contrast, a similar plot of the $a_{1g}$ state of the SQS-PM FeO below the Fermi level shows an interesting spatial disorder in addition to the spin disorder, demonstrated in Figure 3(b). As mentioned in the analysis of the AFM phase, the $a_{1g}$ state in FeO is a singly occupied band resulting from the linear combination of three near-degenerate $t_{2g}$ orbitals. The spatial disorder illustrated in Figure 3(b) reflects the occupancy symmetry breaking coupled to the spin disorder enabled by the SQS model and captured by SCAN, consistent with the findings from DFT+$U$ (see Figure 6 of Ref. 12). The spatial disorder also presents in the AFM FeO when a supercell containing more than two Fe atoms is used (see Appendix B).

In general, the DOS curves of SQS-PM model are similar to those of the AFM model (Figure 2), which is consistent with the experimental finding, i.e., the long-range ordering of the magnetic moments is not the driving force of the band gap opening. Instead, it is the stabilization of the *local* magnetic moment that opens the band gap, which is the essence of the Mott physics,[1] consistent with previous studies of local spin density approximation (LSDA) plus self-interaction correction (SIC)[44] with disordered local moment (DLM)[30] implemented within the single-site coherent-potential approximation (CPA)[45,46] and DFT+DMFT results.[7,8]

### D. Symmetry Restoration

As in all symmetry broken solutions[15,47,16] one needs to restore the global symmetry, see Ref. 47 and references therein. It is telling that different SQS configurations corresponding to different initial nudging all give similarly large local moments and similar band gaps.[12] Additionally, as Ref 12 noted, whereas different initializations of SQS create slightly different total energies and band gaps, the coupling between such broken symmetry solutions needed to restore symmetry would be very weak as they are spatially localized in different subspaces. We therefore expect that symmetry restoration will have little influence on the total energy and band gaps but will generate a symmetrized wavefunction.

Approaches exist for restoring symmetries after they are broken, e.g., projected unrestricted HF and coupled-cluster-singles-doubles (CCSD) in quantum chemistry,[15,47] and in nuclear physics.[16] As spontaneous symmetry breaking happens in extended systems however, e.g., the AFM phase of the monoxides considered, the symmetry dilemma of getting total energy and wavefunction symmetry simultaneously correct may be less severe for solids. Special care must be paid to properties determined by symmetries however, e.g., topological properties of materials with open *d*- or *f*-shell ions, when computed from the symmetry-broken DFT calculations, as symmetry restoration is likely required.

## IV. Understanding how exchange correlation functionals and self-interaction correction (SIC) affect band gap opening for correlated materials in conjunction with symmetry breaking representations

There are two practical definitions of the band gap concept. The band gap defined as the "total energy difference band gap" (also termed the "delta SCF method", see Ref [48]) is the separation between the ionization energy $E_I = E(M-1) - E(M)$ and the electron affinity energy $E_A = E(M) - E(M+1)$, where $M$ is the number of electrons. Since in this definition only the ground state energies are involved, this total energy band gap can be calculated, in principle exactly from DFT and directly compared to that measured from experiments.[49-51]

On the other hand, the band gap defined through the "*single particle energy band gap*" $\varepsilon^{CBM} - \varepsilon^{VBM}$ (where CBM and VBM denote conduction band minimum and valence band maximum, respectively) is obtained from a Kohn-Sham (KS) or generalized KS (gKS) DFT calculation. Unlike the gKS, in the KS scheme, the XC effective potential is multiplicative. For example, the generalized-gradient-approximation (GGA) in the form of PBE[39] has a multiplicative effective potential and cannot consistently open the band gap defined through the single particle energies of the four monoxides considered here even in the AFM phase. However, the generalized KS scheme allows the DFT effective potential to be non-multiplicative and continuous (e.g., orbital dependent). It has been proved[49-51] that in this scheme the single-particle $\varepsilon^{CBM} - \varepsilon^{VBM}$ band gap for a solid from a density-functional calculation is equal to the total energy band gap for the same density-functional, if the gKS potential operator is continuous and the density change is delocalized when an electron or hole is added. The above noted proof also implies that in the gKS scheme, density-functional that are improved for giving better total energies can predict improved band gaps for solids. SCAN is orbital dependent and implemented in the gKS scheme with a continuous effective potential and thus not restricted by the first error (failure to distinguish occupied from unoccupied orbitals) discussed below.

There are three major, XC-related conditions for minimizing the underestimation of the band gap predicted from the single particle energies:

### A. Use of XC functionals that distinguish occupied from unoccupied states and the issue of XC derivative discontinuity

An ideal density functional should be nonlocal, which usually is realized by orbital dependence, i.e, distinguishing occupied from unoccupied states. To use orbital energies to predict total energy band gaps, it is important to perform calculations in the gKS scheme to eliminate errors associated with the derivative discontinuity.[49-51] In Refs [12] and [13] the authors emphasized this condition for exchange correlation functionals.

In the KS scheme the used exchange correlation energy functionals, e.g., the orbital-dependent ones, can have a derivative discontinuity that is present even for the exact KS exchange-correlation density functional.[52] The discontinuity is difficult to compute and generally ignored, but it should be added to the single particle band gap for the correct prediction of total energy band gap.[52] When SCAN is implemented in gKS, there is no derivative discontinuity in its effective potential, having an analytic continuous behavior.

### B. Use of XC functionals that minimize SIE, enforced by the piecewise linearity of total energy.

The used exchange-correlation density functionals can suffer from the self-interaction error[44] (SIE was also discussed and explained early on by Zunger and Freeman.[48]) SIE results from the imperfect cancellation of the spurious classical Coulomb self-interaction by the approximate exchange-correlation self interaction. Because the repulsive self-Coulomb exceeds the attractive self-exchange correlation, the net SIE is generally positive, causing orbitals to be under bound (orbital energies too high) and wave functions to be excessively delocalized. Such overly delocalized orbitals may not have the needed spatial resolution to 'see' symmetry breaking and can underestimate the total energy band gap.

The SIE for a single orbital can be easily defined and clearly illustrated in single-electron systems [Figure 4(a)] while the many-orbital SIE has to be defined through the deviation from the piecewise linearity of total energy between two adjacent integers of electron number. The exact density functional should have a linear connection between two adjacent integers with derivative discontinuities at integer $M$,[52] also known as the generalized Koopman's condition. Many popular semilocal density functionals (e.g., PBE) generally over-stabilize the energy for fractional numbers, deviating from the piecewise linearity from below [Figures 4(c) and 4(d)] and indicating the tendency to spuriously delocalize electrons between nuclear centers (i.e., the delocalization error).

One-electron self-interaction correction that is related to the correct power-law in vacuum (i.e. far away from the atom) is not sufficient for improving predictions of band gaps. Correcting the gap due to SIC affects also greater orbital compactness on the length scale of the chemical bond. This is often referred to as reducing the delocalization error stemming from SIE. Indeed, for a DFT calculation in the gKS scheme with symmetry breaking, it is the many-electron SIE or delocalization error that matters for the band gap predictions from orbital-dependent density functionals. The lower the many-electron SIE or delocalization error, the lower the error in band gap predictions.[53,54,50] Many density functionals, which are one-electron SIE free, including weighted density approximation (WDA)[55] and the Perdew-Zunger (PZ) self-interaction correction (SIC),[44] have also reduced many-electron SIE and improved band gap predictions.[56-59,54]

The PZ-SIC[44] or an approximation thereof via the DFT+$U$ approach[12,13,60,40] opens the band gaps of the aforementioned binary and ABO$_3$ 3$d$ oxides by creating a distinction between occupied and unoccupied states while also creating more compact orbitals able to take advantage of symmetry breaking possibilities. For the binary 3d oxides considered here, SIC has been shown to improve the band gap prediction of LSDA dramatically by ~1-3 eV,[56] while DFT+U with modest U that has the primary effect of reducing SIE similarly increases band gaps from zero to 1 eV or more in 3d oxides.[12,13]

Most widely used density functionals did not implement the condition of self-interaction cancelation into their constructions. Yet, any functional can be measured as to the extent of its *de facto* self-interaction cancellation, even if this condition was not deliberately implemented. We show in Figures 4(c) and 4(d) that SCAN has a better self-interaction cancelation than PBE does.

Not all mean-field approximations are alike. The Hartree-Fock (HF) approximation is orbital-dependent, and one-electron exact and thus self-interaction free, (so it satisfies the first two conditions listed in the current Sec. IV), and satisfies the Koopman's theorem which dictates the first ionization energy to be equal to the negative of the orbital energy of the highest occupied molecular orbital under the frozen orbital approximation. HF however misses all correlation energy and the effective screening of the exchange interaction. It thus ends up being a very poor approximation, badly overestimating band gaps. This overestimation is a consequence of the over-localization error in correspondence to the deviation of HF from the piecewise linearity in total energy between two adjacent integer numbers (see Figure 4(c) and 4(d)). This suggests that the orbital-dependence condition is only a necessary condition for a density functional which should also satisfy the generalized Koopman's condition for accurate predictions of the total energy band gap using its orbital energies. For conventional density functionals, satisfying the generalized Koopman's condition enforces the minimization of SIE.

### C. Use of XC functional with spin symmetry breaking

If one insists on no symmetry breaking, then for strongly correlated electron systems, including the $3d$ oxides, the exchange-correlation energy functional should have an *explicitly* discontinuous (non-differentiable) dependence on the density or density matrix. This traditional no symmetry breaking scenario is illustrated by the NM phases of the binaries as well as by the spin singlet $H_2$ molecule (compared to the spin symmetry broken solution of $H_2$ where the electron at one proton is spin up and the other electron spin down at the other proton, reminding of AFM against NM).[61] Such discontinuity is missing from all current practical exchange correlation approximations. This missing piece results in too high energy as seen in the NM phases of the binaries, and should be added to the orbital energy band gap for the prediction of the total energy band gap. This source of error is removed if the spin symmetry is allowed to break as in our present approach as well as refs 12-14. Note that even LDA/GGAs accommodate spin symmetry breaking. However, the problem for LDA/GGAs is that these functionals can not take advantage of this spin-symmetry breaking given their internal restrictions of not using conditions 1 (orbital-dependence) and not using condition 2 (SIE reduction). In other words, our 3 conditions must be applied simultaneously, not in isolation.

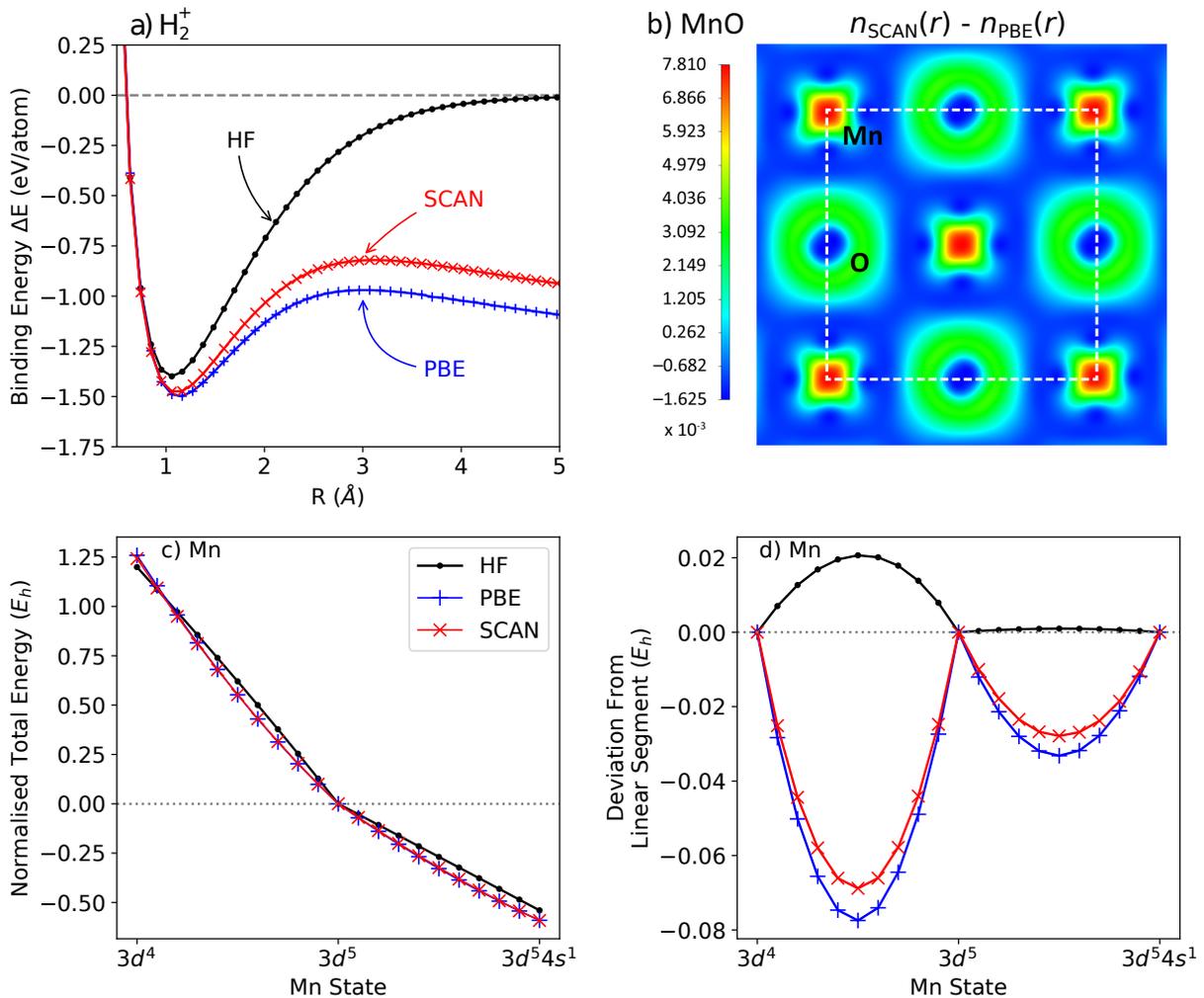

**Figure 4.** Self-interaction error reduction and electron localization with SCAN. (a) The dissociation energy curves of the $H_2^+$ molecule from Hartree-Fock theory (HF), PBE, and SCAN. HF theory is exact for the chosen basis set for single electron $H_2^+$. (b) Difference of electron density calculated by SCAN and PBE, i.e., $n_{SCAN} - n_{PBE}$. AFM MnO is selected because $Mn^{2+}$ is fully spin polarized with a simple completely filled $3d^5$-shell in one spin channel. (c) Total energy of isolated Mn ion as a function of fractional orbital occupation, normalized such that the $3d^5$ ($Mn^{2+}$) state is at zero. (d) Deviation from ideal linear behavior between integer electron numbers for the total energies of sub-figure (c).

The SCAN functional systematically opens qualitatively correct band gaps without invoking the $U$ parameter (Table I). This is understood by the following three factors corresponding to the three error sources mentioned above: (i) SCAN is orbital-dependent and implemented in the *generalized* Kohn-Sham (gKS) scheme; (ii) SIE is reduced in SCAN compared to PBE; (iii) Breaking of the spin-symmetry eliminates the third error source above and also facilitates accurate total energy descriptions for these monoxides in their AFM and PM phases in comparison to their NM phases.

## V. Illustration of smaller SIE and greater orbital compactness in SCAN: The hydrogen molecule and hydrogen chain

The one-electron SIE can be demonstrated in the $H_2^+$ binding energy curve [Figure 4(a)]. As there is only one electron, the Hartree-Fock (HF) description is exact because the classical Coulomb interaction is completely canceled by the exact exchange.[62,63] When stretching $H_2^+$, significant SIE develops within the PBE and SCAN calculations, while SCAN reduces the SIE noticeably. Figure 4(c, d) show the total energy *vs.* fractional electron number ($E$ *vs*. $N$) of Mn ion and the deviations from the linearity of total energy between adjacent integers, respectively. Consistent with the performance in $H_2^+$, SCAN reduces the deviation noticeably and thus the delocalization error or many-electron SIE compared with PBE.

For SCAN, the SIE reduction leads to more compact and energetically deeper orbitals, as directly visualized in the electron density difference, $n_{\text{SCAN}} - n_{\text{PBE}}$, plot for MnO [Figure 4(b)]. The charge density difference shows that SCAN accumulates electrons around the ions but depletes them in the interstitial regions compared to PBE, a clear evidence of electron localization and SIE reduction.

The ability of SCAN to capture a part of SIC helps it describe nontrivial two electron systems, e.g. $H_2$, with accuracy rivaling the far more complex coupled cluster singles and doubles method that is exact in two electron systems for the chosen basis set. Figure 5(a) shows the $H_2$ binding curves, which illustrate the success and limitations of the spin-symmetry breaking approach for modelling the PM phase. As the $H_2$ bond length is stretched, the first excited spin-triplet and ground spin-singlet states become degenerate, causing difficulty for mean-field-like methods if symmetries are enforced. If spin-symmetry is broken even a spin unrestricted HF model can recover the exact dissociation limit, though it deviates from the reference curve at shorter separations. Remarkably, unrestricted SCAN model with the spin-symmetry breaking accurately recovers the whole reference binding curve with only small disagreement around the shoulder (at 1.5 Å) to a maximum of ~0.1 eV/atom. The recovery of the binding energy curve comes at the price of "spin contaminated" unrestricted single-particle KS wavefunction solutions however, which are no longer eigenfunctions of the total spin operator $\hat{S}^2$.

Spin-symmetry broken SCAN performs similarly well for the 10-hydrogen chain [Figure 5(b)], and the thermodynamic limit of the infinite hydrogen chain $H_\infty$ [Figure 5(c)]. A small increase in error around equilibrium is seen in these cases. Being regarded as a generalization of the one-dimensional Hubbard model, $H_\infty$ captures the key features that include the presence of strong electron correlation of diverse nature as the H-H distance is varied, and a need to treat the full physical Coulomb interaction and to work in the continuous space and thermodynamic limits.[31] To our knowledge, $H_\infty$ is the only extended correlated system with accurate reference total energies that can be used to benchmark density-functionals for the strong correlation.

It is striking that a density-functional single determinant[64] produces such accurate energies when the spin-symmetry is allowed to break. Note that spin-symmetry breaking solutions from DFT+SIC have also been shown to be critical for describing response properties of hydrogen chains,[65] consistent with our supposition that Koopman's compliance of the functional [Figures 4(c,d)] helps in approaching the correct ground state.

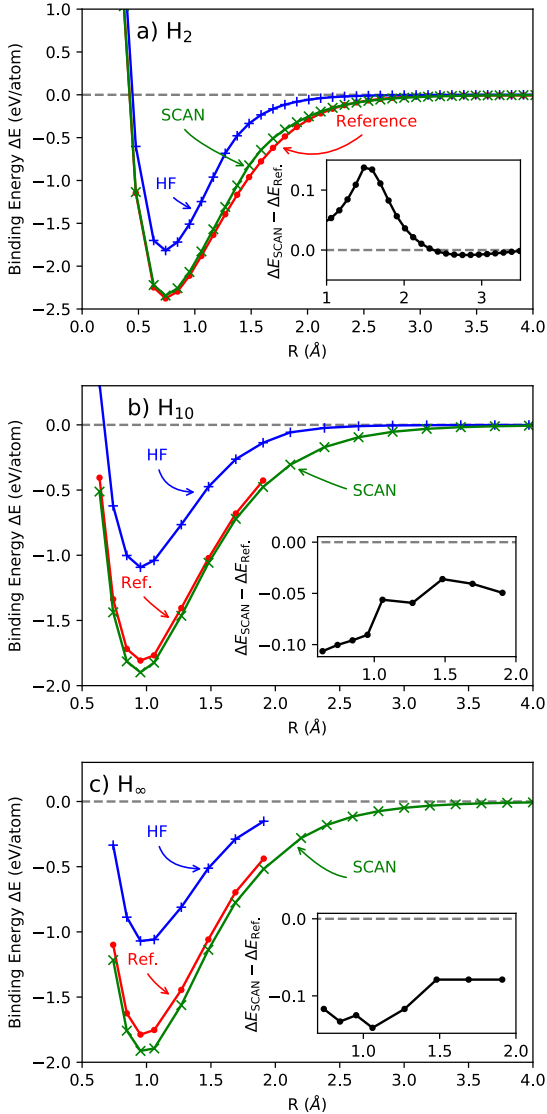

**Figure 5.** Performance of unrestricted SCAN approach with spin-symmetry breaking for various hydrogen structures. Binding energies are plotted as a function of interatomic distances for (a) $H_2$, (b) $H_{10}$, and (c) the thermodynamic limit of the hydrogen chain ($H_\infty$). In (a), the coupled-cluster-singles-doubles (CCSD) (red ●) is exact and used as the reference. In (b) and (c), the HF data < 1.9Å (blue +) and reference auxiliary-field quantum Monte Carlo data (red ●) are from Ref 31. $H_\infty$ has an anti-ferromagnetic order when the magnetic moment persists. Atomic spacing is uniform in $H_{10}$ and $H_\infty$ systems, the same value as used in Ref. 31. The insets show the energy difference between SCAN and the references.

As with the binding curves of these hydrogen systems, it is reasonable to expect that spin-symmetry broken SCAN with the SQS model can predict better total energies for the PM phase than the spin-symmetry conserved NM model. With the connection between total energy and band gap established in gKS[49-51], opening a band gap in the PM monoxides within the SQS model is also expected.

## VI. The role of polymorphous representation

In principle, a density functional with the explicit derivative discontinuities that account for spin degeneracies[61] might predict band gaps in the NM model. However, such a density functional that is also generally applicable to different systems is extremely hard to construct and not available currently. It is important thus to realize that using any of the current exchange correlation functionals without a symmetry breaking polymorphous representation does not open band gaps uniformly in all binary oxides considered here and $ABO_3$ ternaries. This hence mandates the polymorphous representation to allow symmetry breaking (in spin local environments as well as in positional relaxation) for band gap predictions with exchange correlation functional that satisfy conditions discussed in section IV.

Another example where the allowance of different local environments is sanctioned by DFT and provides a qualitative correction of the electronic structure is bond disproportionation.[66] Large, energy-lowering atomic relaxation can convert crystal structures that have a monomorphous structure with a "single local environment" (SLE, such as a single $BX_6$ octahedron in $ABX_3$ perovskites) to a doubled unit cell that is characterized by a "double local environments" (DLE, such as two $BX_6$ octahedra per cell, one small and the other large). For transition metal compounds, correlation effects were previously argued to be the reason for such transitions.[67] However, from standard DFT calculations for both s-p electron systems such as $BaBiO_3$ and $CsTlF_3$ as well as for d electron systems such as SmNiO$_3$ and CaFeO$_3$, it was recently found[66] that whenever the SLE phase is metallic, the formation of the DLE polymorphous configuration lowers the total energy and becomes automatically insulating, in agreement with experiment. Thus, the metal-insulator transition in these systems is a consequence of structural symmetry breaking that is systematically captured by DFT for a broad range of either *sp* or *d*–electron compound without a need for special effects.

The polymorphous situation can encompass local variations in (i) atomic displacements in AFM and PM including Jahn-Teller distortions,[68] octahedral tilting and rotation (ii) local magnetic moments in AFM and PM (iii) spin local environments in the PM phases, and (iv) distribution in occupation patterns of partially occupied degenerate partner levels. Unlike the Mott-Hubbard approach that explains band gap opening by a uniform mechanism ("*U*") for all Mott insulators alike, the polymorphous approach finds that gapping in different members of the binary and ternary 3*d* oxides series are dominated by different local effects from the list (i)-(iv) above, either using DFT+$U$[12,13,68] or DFT-without *U* (ref 14 and the present work) alike. The polymorphous approach explains not only band gap opening in these compounds, but also orbital ordering[14] and reveals quantitative agreement with measured local moments[12-14] and atomic displacements.[68]

The significance of the polymorphous representation on gapping is clear: Naïve DFT approaches have often modeled the properties of the system as the property <P>=P(S$_0$) of the macroscopically averaged monomorphous structure S$_0$ rather than the average P$_{obs}$=ΣP(S$_i$) of the properties {P(S$_i$)} of the individual, low symmetry microscopic configurations {S$_i$ ; i=1, N}. The SQS provides a direct route to P$_{obs}$: The observable property P (*e.g.*, band gap) calculated for an SQS structure is more than the average property from many small random structures, but approximates the ensemble average P for the polymorphous configuration.[20,22] It has been shown that a relatively small SQS structure produces numerically the same property values as well as larger (ergodic) randomly selected supercells do.[69]

## VII. Conclusions

This work shows that the non-empirical and efficient semilocal SCAN density-functional in conjunction with the polymorphous representations predicts reliable magnetic moments and opens band gaps for the prototypical Mott 3*d* transition-metal monoxides in the AFM and PM phases, even without interelectronic *U*.[12] We have thus demonstrated that mean-field-like theories can describe critical properties like band gaps and magnetic moments of open *d*- and *f*-shell materials when symmetry breaking is allowed. This work therefore opens an alternative to symmetry-conserving approaches for the study of open *d*-shell and *f*-shell materials. The use of such non-naïve DFT for applications to complex 3*d* systems where other approaches may be computationally prohibitive (catalysis, photovoltaics, and multi-component device structures) is therefore legitimate.

This work identifies, out of many possible other combinations, 4 specific conditions, that when used together, go a long way to solve the band gap opening problem encountered with naive DFT applications to Mott insulators. The first 3 conditions apply to the exchange correlation functional, whereas the fourth condition applies to the real-space representation of the structure.
(i) Use of XC functionals that distinguish occupied from unoccupied states.
(ii) Use XC functionals that minimize self-interaction error, enforced by the piecewise linearity of total energy.
(iii) Use of XC functionals with spin symmetry breaking.
(iv) Allow unit cell representation that could break symmetry, possibly leading to polymorphous networks. The latter can encompass local variations in quantities such as atomic displacements, local magnetic moments, local spin environments, and distribution in occupation patterns of partially occupied degenerate partner levels.
It is important to emphasize that whereas individual conditions, when considered in isolation, may not be sufficient, when taken together they are both sufficient and necessary (although quantitatively not perfect). The Hartree-Fock approximation misses all the correlation effect, and doesn't fall into the categories to be considered here although it satisfies the above requirements for the XC functionals.

**Acknowledgments**

The work at Tulane was supported by the U.S. DOE, Office of Science, Basic Energy Sciences grant number DE-SC0019350. Calculations were partially done using the National Energy Research Scientific Computing Center supercomputing center and the Cypress cluster at Tulane University. Y.Z. thanks Hoang Tran, Carl Baribault, and Hideki Fujioka for their computational support. The work at the University of Colorado at Boulder was supported by the National Science foundation NSF Grant NSF-DMR-CMMT No. DMR-1724791 and XSEDE computer time. J.S. and A.Z acknowledge fruitful discussions with Gustavo Scuseria, Mark Pederson, John Perdew, and Samuel Trickey. JS acknowledges also discussions with Weitao Yang, Christopher Lane, Bernardo Barllieni, Robert Markiewicz, and Arun Bansil. We are grateful to G.Trimarchi for continued important discussions.

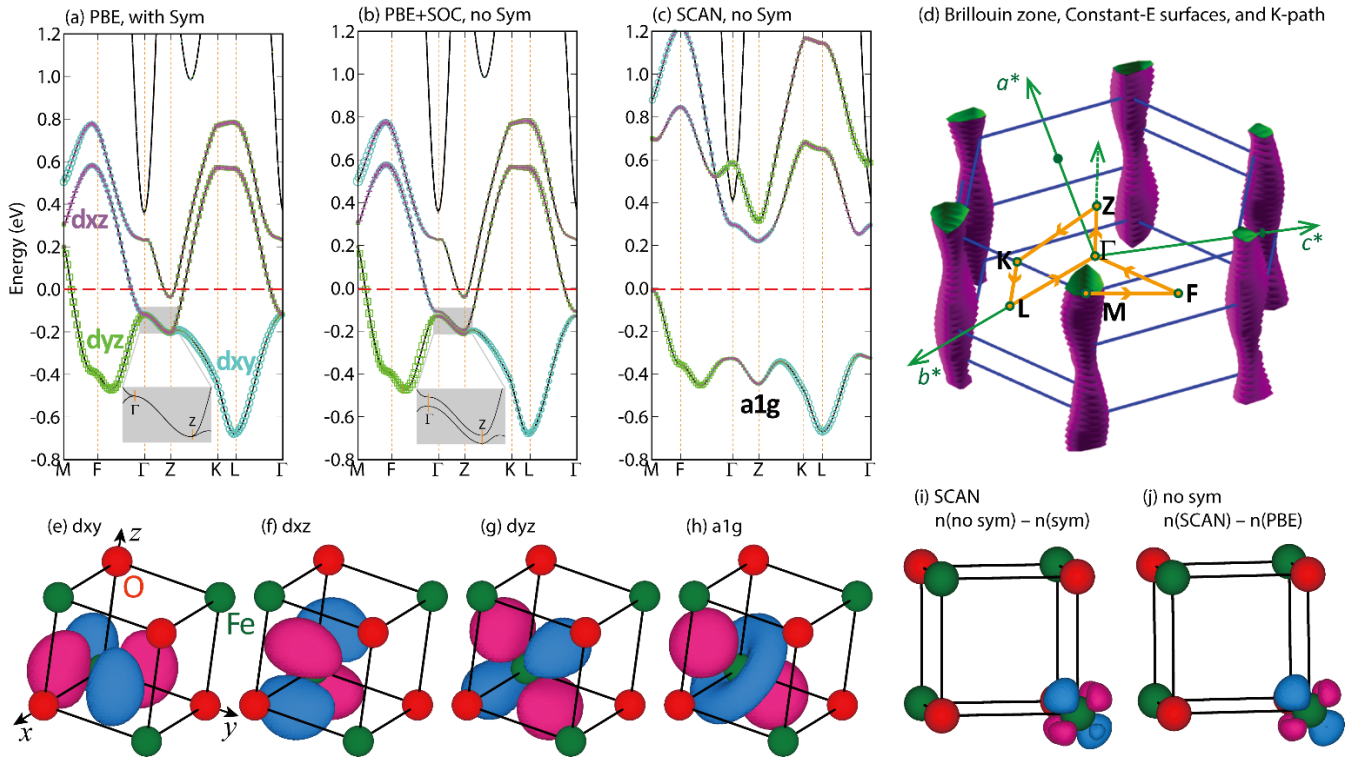

**Figure 6.** Gap opening mechanism in rock-salt FeO with the G-AFM magnetic ordering. (a) PBE predicted band structure by keeping the $t_{2g}$ orbital symmetry. (b) PBE band structure but with the $t_{2g}$ orbital symmetry removed. Spin-orbit coupling effect is used to guide the calculation to find the electronic ground state (Zhi Wang: what does it mean? Are we using SOC in this PBE calculations or not? Or in a special way?). (c) SCAN band structure also with orbital symmetry removed. In subplots (a-c), the $t_{2g}$ wavefunctions are superimposed onto the band structures. The insets of subplots (a-b) show the band degeneracies of the zoomed-in areas. (d) Brillouin zone and K-path of the G-AFM phase for the band structure calculations, together with the energy surfaces that are 0.1 eV below the valence band maximum in subplot (c). (e-g) The $d_{xy}$, $d_{yz}$, $d_{xz}$ orbitals extracted from the Wannier function construction, from SCAN calculation. The blue and pink colors denote the signs of Wannier function. (h) The $a_{1g}$ orbital. (i) The difference of charge density distribution between calculations with and without orbital symmetry, both from SCAN calculations. The blue and pink colors denote charge accumulation and depletion, respectively. (j) The difference of charge density distribution between SCAN and PBE calculations, both without orbital symmetry.

# APPENDIX A: Opening a Band gap in AFM FeO and CoO: Occupation Symmetry Breaking of the $t_{2g}$ Orbitals

$Fe^{2+}$ ($3d^6$) in FeO and $Co^{2+}$ ($3d^7$) in CoO have their majority spin channels filled by five electrons, but their three-fold degenerate $t_{2g}$ orbitals in the minority spin channels have partial occupancy with one and two electrons, respectively. Therefore, the degeneracy of the three $t_{2g}$ orbitals must be lifted in order to open a band gap across them. Figure 1(c,d) show that the band gaps from SCAN are between one occupied $t_{2g}$ band and two empty $t_{2g}$ bands for FeO, two occupied $t_{2g}$ bands and one empty $t_{2g}$ band for CoO. Although the gap opening mechanism is similar for FeO and CoO, we find that it is much more difficult to open the band gap of FeO. We, therefore, take FeO as an example to analyze the orbital physics and gap opening mechanism.

Figure 6(a-c) are the $t_{2g}$ bands of FeO calculated by density-functional approximations. When orbital symmetry (see detailed discussions in the next paragraph) is enforced, a common method for reducing computational cost, neither PBE [Figure 6(a)], SCAN (not shown here), nor their combination with Hubbard $U$ opens a gap across the three $t_{2g}$ orbitals. The degeneracy of the three $t_{2g}$-derived bands along the Γ-Z path is protected by the orbital symmetry, which prevents a gap being opened within the single-particle approaches. After lifting the orbital symmetry constraint, PBE opens a tiny local gap at each K-point [inset of Figure 6(b)], although those local gaps are too small to result in a band gap across the whole Brillouin zone. On the other hand, the SCAN meta-GGA separates one occupied $t_{2g}$ orbital [the $a_{1g}$ orbital, see Figure 6(c)] from the other two unoccupied $t_{2g}$ orbitals, resulting in an insulating state.

The gap opening in SCAN calculations is partially due to the SIE reduction. The SIE typically makes the $pd$ orbitals too diffuse, leading to a too small band gap. Another important effect is related to the orbital occupation symmetry breaking: SCAN predicts that the $d_{xy}$, $d_{yz}$, and $d_{xz}$ [Figure 6(e-g)] orbitals have different occupation numbers of 0.332, 0.342, and 0.272 electrons, respectively. It is worth noting that there are some numerical uncertainties in counting the electron numbers based on the DFT approaches. Nevertheless, it is evident that while the $d_{xy}$ and $d_{yz}$ orbitals are near-degenerate, the $d_{xz}$ orbital has a smaller occupation.

This $t_{2g}$ orbital polarization can only be captured when the constraint of orbital symmetry was removed, which can be realized by turning on spin-orbit coupling or turning off the symmetry constraint (ISYM = −1 in the VASP calculation). The linear combination of the $d_{xy}$, $d_{yz}$, and $d_{xz}$ orbitals has $a_{1g}$ symmetry [Figure 6(h)], and its orientation is approximately along [111] direction with a small deviation angle. As a result, the orbital symmetry is lower than the rock-salt lattice symmetry. In fact, the $t_{2g}$ orbital polarization can be directly visualized from the difference of charge densities calculated with and without the symmetry constraint [Figure 6(i)], which reveals an $a_{1g}$-like shape with charge accumulation along the [111] direction.

**Table II.** Relative stability of FeO from different simulation models. "Un-relaxed" means that both the lattices and internal coordinates are kept to the experimental NaCl crystal structure, while "relaxed" means that all structural degrees-of-freedom are fully relaxed. Note that the cubic $FeO_6$ octahedron is distorted in the structural relaxation. For the SQS-PM phase, our simulations with the 64-atom supercell (not shown here) are qualitatively similar to the 216-atom supercell results here. The bigger (216-atom) cell is used here to better represent the spin disordering effect. The cubic 216-atom supercell is, however, incommensurate with the G-type AFM order.

| Simulation models | | Spin long-range ordered? | Orbital long-range ordered? | $FeO_6$ cubic symmetry? | Energy (meV/atom) |
|---|---|---|---|---|---|
| Un-relaxed | Primitive cell (4 atoms) with G-type AFM | Yes | Yes | Yes | 0 |
| | Supercell (64 atoms) with G-type AFM | Yes | No | Yes | -31 |
| | Supercell (216 atoms) with SQS-PM | No | No | Yes | -18 |
| Relaxed | Primitive cell (4 atoms) with G-type AFM | Yes | Yes | No | -59 |
| | Supercell (64 atoms) with G-type AFM | Yes | No | No | -90 |
| | Supercell (216 atoms) with SQS-PM | No | No | No | -63 |

Theoretically, it is interesting that SCAN captures the $t_{2g}$ orbital polarization well. For comparison, we plot the difference of charge distribution predicted by SCAN and PBE [Figure 6(j)], both without orbital symmetry constraint in the calculations. Surprisingly, this pattern in Figure 6(j) is almost identical to that in Figure 6(i). As PBE usually underestimates the electron inhomogeneity, our results show that SCAN is more reliable than PBE in recognizing the subtle differences of chemical environments and thus effectively distinguishing the transition-metal $3d$ orbital anisotropy.

**APPENDIX B: Supercell Effects on the FeO Structure Relaxation at the AFM and PM Phases**

Table II shows the effect of supercell size on the FeO structure relaxation. The well-known G-type AFM phase is conventionally simulated in a hexagonal primitive cell with two Fe atoms and two oxygen atoms. In this work, we also simulated the spin-disordered phase using the SQS paramagnetic (SQS-PM) model with 64- and 256-atoms. It is surprising that the FeO SQS-PM phase have lower energies compared with the above AFM phase simulated with 4-atom cell. This is not consistent with those in MnO, NiO, and CoO (see Table I).

The reason for the energy of the 216-atom SQS-PM being lower than the 4-atom AFM of FeO calculations can be understood by considering the effect of an additional degree of freedom, the orbital spatial order [see Figure 3(b)]. The orbital is enforced to order in the 4-atom primitive cell with the G-type AFM configuration, but with larger supercells the long-range orbital ordering can be broken. We therefore simulate the two effects, i.e., spin ordering and orbital disordering, for the AFM phase using a 64-atom supercell (Table II). Interestingly, this model has the lowest energy among all three simulation models shown in Table II, indicating the critical role of orbital disorder in stabilizing the electronic energy. To directly visualize these effects, we also plot the charge density of the $a_{1g}$ orbital in Figure 7, similar to Figure 3.

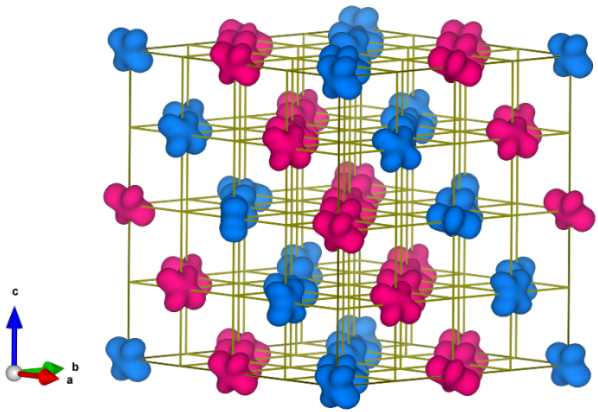

**Figure 7.** Charge density of the $a_{1g}$ orbital of FeO in the 64-atom supercell. The G-type spin configuration is shown as FM coupling within (111) plane and AFM coupling between the planes. The $a_{1g}$ band shows a disordered pattern in charge density. Oxygen atoms are omitted for simplicity.